\documentclass[epsfig]{mn2e}                   
\usepackage[dvips]{graphicx}
\usepackage[dvipdfm,CJKbookmarks,bookmarks]{hyperref}
\newcommand{\Msolar}{\mbox{\,$\rm M_{\odot}$}}        

\begin{document}
\title[Black hole masses in NLS1s]{Black hole masses in narrow-line Seyfert 1
galaxies}
\author[Bian and Zhao]{W. Bian$^{1,2}$\thanks{E-mail: whbian@njnu.edu.cn} and
Y.Zhao$^{2}$\\
$^{1}$Department of Physics, Nanjing Normal University, Nanjing
210097, China\\
$^{2}$National Astronomical Observatories, Chinese Academy of
Sciences, Beijing 100012, China}
\date{}
\maketitle
\begin{abstract}
The masses of central supermassive black holes in a soft X-ray
selected sample of the narrow-line Seyfert 1 galaxies (NLS1s) are
estimated by some different methods to test their theoretical
models. Apart from the methods using the H$\beta$ linewidth and
the [O III] linewidth, soft X-ray excess as a prominent character
of NLS1s is used to estimate the black hole masses. The virial
mass derived from the H$\beta$ linewidth assuming random orbits of
broad-line reigns (BLRs) is consistent with that from the soft
X-ray bump luminosity for NLS1s but with a larger scatter. The
virial black hole masses showed that most of NLS1s are in the
super-Eddington accretion state while most of broad-line Seyfert 1
galaxies (BLS1s) are not. We found that the black hole mass
estimated from [O III] linewidth is not in agreement with above
two methods. Using the Eddington limit relation for the
super-Eddington accretion suggested by Wang (2004), we found that
there are 16 NLS1s satisfied with this Eddington limit relation.
The masses of these 16 NLS1s derived from X-ray luminosity are
systematically larger than that from H$\beta$ linewidth assuming
random BLRs orbits. If the mass derived from X-ray luminosity is
true, the mean disk inclination to the line of sight in these 16
NLS1s is about $17^{\circ}$, which provided new support for the
pole-on orientation effect in NLS1s.

\end{abstract}

\section{INTRODUCTION}

Narrow-line Seyfert 1 galaxies (NLS1s) are a peculiar class of
active galactic nuclei (AGNs). They are characterized (Osterbrock
\& Pogge 1985): smaller H$\beta$ FWHM (less than $2000km~s^{-1}$),
strong optical Fe II multiplets, the line luminosity ratio of
[O~III] 5007$\AA$ to H$\beta$ is less than 3, steep soft X-ray
excess (Boller et al. 1996), and rapid soft/hard X-ray variability
(Leighly 1999; Cheng et al. 2002). A popular model of NLS1 is that
they contain less massive black holes, but have higher accretion
rates radiating at close Eddington luminosity, namely high
Eddington ratios (Pounds et al. 1995; Laor et al. 1997; Mineshige
et al. 2000). It has been suggested that NLS1s might be in the
early stage of AGNs evolution (Grupe 1996, Grupe et al. 1999,
Mathur 2000; Bian \& Zhao 2003a) and black hole grows fast via a
higher fraction of Eddington accretion rate (Sulentic et al. 2000;
Boroson 2002). To test this hypothesis we need to estimate the
black hole mass of NLS1s.

There are several methods to calculate the black hole masses in
AGNs: 1) virial mass derived from the H$\beta$ FWHM and the sizes
of broad line regions (BLRs) from the reverberation mapping
technique or the empirical size-luminosity formula (Ho 1998;
Wandel et al. 1999; Kaspi et al. 2000; Peterson et al. (2000);
Vestergaard 2002; Bian \& Zhao 2003b; 2004); 2) soft X-ray
variability (Czerny et al. 2001; Bian \& Zhao 2003c); 3) the
relation between the mass and the bulge stellar velocity
dispersion (M-$\sigma$ relation) or the bulge luminosity
(M-$L_{bulge}$ relation)(McLure \& Dunlop 2001; Tremaine 2002). We
are not sure whether above methods can apply to NLS1s for their
special properties. Applying reverberation mapping technique to
NLS1s is difficult since they are usually less variable in optical
band (Shemmer \& Netzer 2000). Whether the assumption of the
random BLRs orbits in NLS1s is suitable or not is a question open
to debate (Bian \& Zhao 2002). The empirical size-luminosity
relation is only based on many broad-line Seyfert 1 galaxies
(BLS1s) and a few NLS1s (Kaspi et al. 2000). Peterson et al.
(2000) suggested that NLS1s and BLS1s follow the same
size-luminosity relation. However, BLRs physics in NLS1s is
possibly special and different compared with BLS1s (e.g. Boller et
al. 1996). Therefore, whether NLS1s follow this relation or not
should be confirmed by future observation in NLS1s. The $M-\sigma$
relation may not apply to NLS1s since they are most likely in the
early stage of AGNs evolution, which is intimately related with
the black hole growth process (Mathur et al. 2001; Wandel 2002; Lu
\& Yu 2003; Shields et al. 2003; Bian \& Zhao 2004; Grupe \&
Mathur 2004).

As an extreme feature in NLS1s, Soft X-ray excess may most likely
be caused by high accretion rate in units of Eddington accretion
rate. This character could be used to probe the black hole mass
since the photon trapping effect gives a saturated luminosity,
namely the luminosity is almost independent to the accretion rate
(Wang \& Zhou 1999; Wang et al. 1999; Ohsuga et al. 2002).
Recently Wang \& Netzer (2003) presented a extreme slim disk with
a hot corona to explain the soft X-ray bump in NLS1s and suggested
that soft X-ray humps in NLS1s are natural consequences of
super-Eddington accretion. They found that the hump X-ray
luminosity is weakly dependent on the accretion rate and almost
completed determined by black hole mass in their model:
\begin{equation}
M_{\rm BH}=2.8\times 10^{6}(\frac{L_{\rm SX}}{10^{44}\rm erg~
s^{-1}})\Msolar
\end{equation}
Where $L_{\rm SX}$ is the soft X-ray luminosity in the flat part
of the bump. It provides new method to estimate the black hole
masses in NLS1s with super-Eddington accretion rates.

In this paper, we compared the results from different methods to
estimate the black hole masses in a sample of NLS1s and BLS1s
(Grupe et al. 2004). We tried to find which method is suitable to
estimate the black hole masses in NLS1s and whether the disk
inclinations to the line of sight in NLS1s compared with BLS1s are
small or not. All of the cosmological calculations in this paper
assume $H_{0}=75 \rm {~km ~s^ {-1}~Mpc^{-1}}$, $\Omega_{M}=0.3$,
$\Omega_{\Lambda} = 0.7$.

\section{Sample}
There are many samples suitable for this kind of research. Boller
et al. (1996) presented an optically selected sample of 46 NLS1s
with extremely steep soft X-ray spectra observed with ROSAT.
Verron-Cetty et al. (2001) compiled a sample of 64 NLS1s and
systematically studied their optical spectra. Willams et al.
(2003) presented a sample of 150 NLS1s found within the Sloan
Digital Sky Survey (SDSS) Early Data Release (EDR), which is the
largest sample of NLS1s. Grupe et al. (2004) presented a complete
sample of 110 soft X-ray selected AGNs adopting the criterion of
Hardness ratio less than zero and found about half of them are
NLS1s. In order to estimate the central supermassive black hole
masses in NLS1s, we used different methods including that from
soft X-ray hump luminosity and that from the H$\beta$ width. Here
we used the sample of Grupe et al. (2004) for its completeness.
This sample can also be used to do a comparable study on NLS1s and
BLS1s (Grupe 2004 ).

\section{Methods}
As the first method, we can estimate the virial masses from
H$\beta$ linewidth based on the assumption that the BLRs clouds
are controlled by the central black hole gravitational potential
(Kaspi et al. 2000). If we know the BLRs sizes ($R_{BLR}$) and
BLRs velocity ($v$), we can derived the black hole mass ($M$)
using Newton law, $M=V^{2}R_{BLR}G^{-1}$. The BLRs sizes can be
derived from the reverberation mapping method or the empirical
size-luminosity formula,
\begin{equation}
R_{\rm BLR}=32.9(\frac{\lambda L_{\lambda}(5100
\rm{\AA})}{10^{44}\rm erg~s^{-1}})^{0.7} ~~\rm{lt-days},
\end{equation}
where $\lambda L_{\lambda}(5100 \rm{\AA)}$ is the monochromatic
luminosity at 5100$\rm{\AA}$. Assuming the random BLRs orbits, the
BLRs velocity can be derived from the H$\beta$ linewidth
($v_{FWHM}$),
\begin{equation}
V=(\sqrt{3}/2) v_{\rm FWHM}.
\end{equation}
This method has been discussed by some authors (Wang \& Lu 2001;
Bian \& Zhao 2003b; Bian \& Zhao 2004; Shields et al. 2003;
Boroson 2003).

As the second method, we can derive the mass from the [O~III]
linewidth. There is a relation between the black hole mass,
$M_{BH}$, and the bulge velocity dispersion, $\sigma$, found in
the nearby normal galaxies (Tremaine 2002):
\begin{equation}
M_{\rm BH}=10^{8.13}(\sigma/(200 \rm km s^{-1}))^{4.02}\Msolar ,
\end{equation}
The bulge velocity dispersion can be derived by the [O~III]
linewidth emitting from the narrow line region (NLRs), where
$\sigma=\rm FWHM(\rm [O~III])/2.35$ (Nelson 2000).

As the third method, we can use the soft X-ray hump luminosity to
calculate the back hole masses in AGNs with the Eddington ratio
$L_{\rm bol}/L_{\rm Edd} > 1$ (see equation (1)).
\section{Results and discussion}


\begin{figure}
\centerline{\includegraphics[width=10cm,height=9cm]{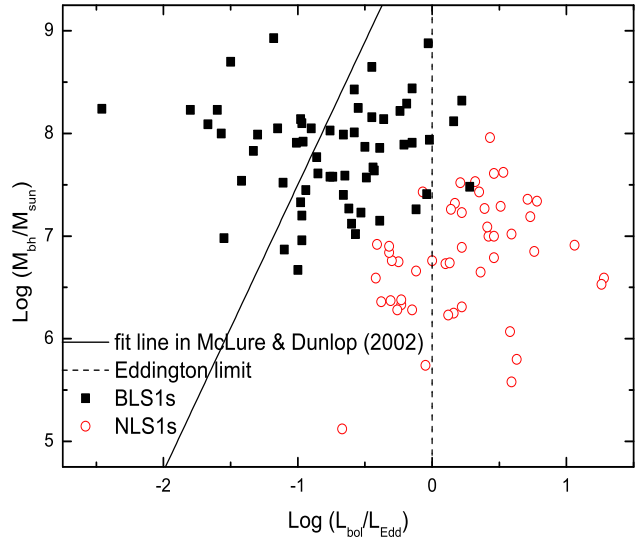}}
\caption{Black hole mass versus bolometric luminosity as a
fraction of the Eddington luminosity. NLS1s are shown as open
circles. BLS1s are shown as solid squares. The location of the
Eddington limit is shown by the vertical dash line. The best-fit
relation for 72 objects found by McLure \& Dunlop (2002) is shown
by the solid line.}
\end{figure}

\begin{figure}
\centerline{\includegraphics[width=10cm,height=9cm]{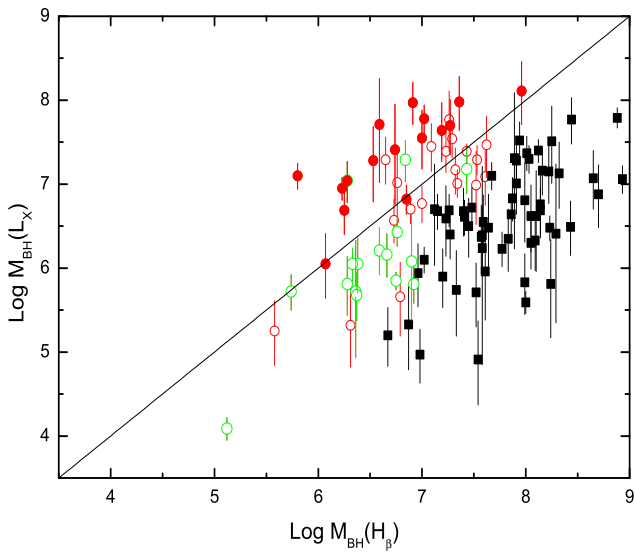}}
\caption{Mass derived from H$\beta$ width versus that from the
soft X-ray luminosity for NLS1s and BLS1s in the sample of Grupe
et al. (2004). NLS1s are shown as circles. BLS1s are shown as
solid squares. The green open circles denote NLS1s with $L_{\rm
bol}/L_{\rm Edd}<1$. The red solid circles denote NLS1s which are
satisfied with the Eddington limit relation defined by Wang
(2004).}
\end{figure}

\begin{figure}
\centerline{\includegraphics[width=10cm,height=9cm]{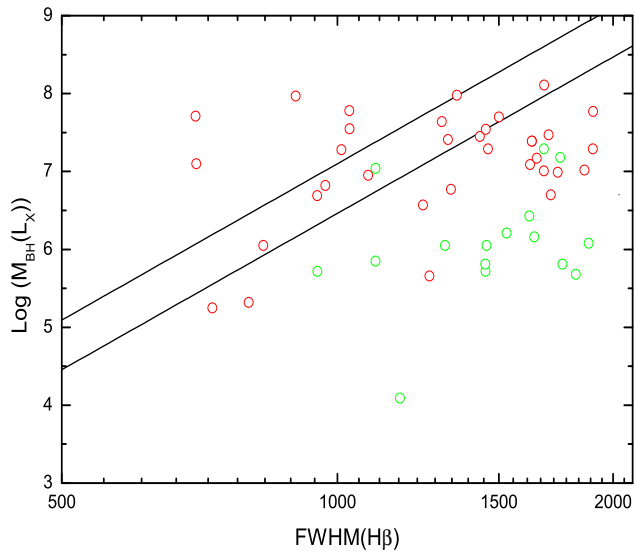}}
\caption{Mass derived from soft X-ray luminosity versus FWHM of
H$\beta$ for NLS1s in the sample of Grupe et al. (2004). Two solid
lines are the Eddington limit defined by Wang (2004). NLS1s are
shown as circles. BLS1s are shown as solid squares. The green open
circles denote NLS1s with $L_{\rm bol}/L_{\rm Edd}<1$.}
\end{figure}

\begin{table*}
\begin{center}
\caption{List of candidates of super-Eddington AGNs using the
limit relation defined by Wang (2004). Col. 1: Name, Col. 2:
Spectral index, Col. 3: Mass from H$\beta$ width in units of
$\Msolar$, Col. 4: Mass from [O~III] width in units of $\Msolar$,
Col. 5: Mass from $L_{\rm SX}$ in units of $\Msolar$, Col. 6: FWHM
of H$\beta$ in units of $\rm km ~s^{-1}$. }
\begin{tabular}{llllll}
\hline
Name & $\alpha_{x}$ & log($M(\rm H\beta)$) & log($M([\rm OIII])$)& log($M(\rm L_{x})$) & FWHM(H$\beta$) \\
\hline
TonS180         &1.89   &6.85   &8.64   &6.82   &970    \\
RXJ0117.5--3826 &2.09   &6.91   &7.93   &7.97   &900    \\
MS0117--28  &2.27   &7.96   &9.39   &8.11   &1681   \\
RXJ0148.3--2758 &2.12   &7.02   &8.83   &7.78   &1030   \\
RXJ0439.6--5311 &2.39   &6.59   &10.01  &7.71   &700    \\
1ES0614--584    &2.46   &6.23   &6.64   &6.95   &1080   \\
RXJ1034.6+3938  &2.38   &5.8    &7.56   &7.1    &701    \\
RXJ1209.8+3217  &3.18   &6.74   &9.1    &7.41   &1320   \\
Mkn766          &1.77   &6.28   &7.16   &7.04   &1100   \\
CBS150          &2.13   &7.36   &8.09   &7.98   &1350   \\
PG1244+026  &1.79   &6.07   &7.51   &6.05   &830    \\
RXJ1304.2+0205  &2.38   &7.19   &9.1    &7.64   &1300   \\
RXJ1319.9+5235  &1.6    &6.25   &6.8    &6.69   &950    \\
QSO1421--0013   &1.72   &7.27   &8.64   &7.7    &1500   \\
RXJ2317.8--4422 &2.87   &6.53   &7.51   &7.28   &1010   \\
MS23409--1511   &2.03   &7  &8.8    &7.55   &1031   \\

\hline
\end{tabular}

\end{center}
\end{table*}

\subsection{Eddington ratio}
For all AGNs in the sample of Grupe et al. (2004) the black hole
virial masses derived from H$\beta$ are firstly calculated. We
also calculated the bolometric luminosity as a fraction of the
Eddington luminosity, $L_{\rm bol}/L_{\rm Edd}$. Here the
bolometric luminosity is from the Table 3 in Grupe et al. (2004),
which is estimated from a combined powerlaw model fit with
exponential cutoff to the optical/UV data and a power law with
neutral absorption to the soft X-ray data (See Fig. 2 in Grupe et
al. (2004)). The bolometric luminosities given in Grupe et al.
(2004) are only approximate, because the EUV part of the spectral
energy distribution of AGN is unobservable and therefore uncertain
(e.g. Elvis et al. 1994). $L_{\rm Edd}$ is derived from the virial
mass via H$\beta$ linewidth. We simply defined the super-Eddington
accretion as $L_{bol}/L_{Edd}>1$. From Fig. 1 we found that for
most of NLS1s $L_{bol}/L_{Edd}$ is larger than one and for most of
BLS1s $L_{bol}/L_{Edd}$ is less than one (also see Fig. 12a and
13a in Grupe 2004). The mean value of $log(L_{\rm bol}/L_{\rm
Edd})$ for 50 NLS1s is $0.23\pm 0.06$ with a standard deviation of
0.44 and for 60 BLS1s it is $-0.75\pm 0.07$ with a standard
deviation of 0.54. Wang \& Netzer (2003) found that for the sample
of Verron-Cetty et al. (2001) the mean value of $log(L_{\rm
bol}/L_{\rm Edd})$ is 0.08. The results from these two samples on
NLS1s are consistent. Most of NLS1s are in the super-Eddington
accretion state while BLS1s are not. The solid line in Fig. 1 is
the best-fit relation for 72 objects found by McLure \& Dunlop
(2002). Although BLS1s seemed to follow the relation found by
McLure \& Dunlop (2002), it is obvious that NLS1s deviated from
this relation, which confirmed our previous results on a small
sample of NLS1s (Bian \& Zhao 2003a). Because the distribution of
the luminosities of BLS1s and NLS1s are very similar (Grupe et al.
2004), NLS1s have to accrete at higher Eddington ratios for their
smaller black hole masses than BLS1s.

\subsection{Black hole masses}
Apart from using equations (2)-(3) to calculate the black hole
masses in AGNs, we also used equation (1) to calculate the masses
in AGNs with super-Eddington accretion. In order to do a contrast,
we also used equation (1) to calculate the mass of AGNs with
sub-Eddington accretion in the sample of Grupe et al. (2004). The
soft X-ray flux, the spectral index $\alpha$ ($F_{\nu} \propto \nu
^{-\alpha}$) and its error are from Grupe et al. (2001). We
calculated the soft X-ray integrated luminosity in 0.2-2.0keV
using the flux and the redshift. Using the spectral index $\alpha$
and the soft X-ray integrated luminosity, we calculated the
monochromatic luminosity at 0.1keV as the luminosity at the flat
part of bump, which would give the black hole mass from equation
(1). Here we considered the error of the spectral index $\alpha$
to calculate the error of the luminosity at 0.1keV and then the
error of mass using equation (1). The mean error in NLs1s is about
0.4 dex. In Fig. 2 we plotted the mass derived from the soft X-ray
luminosity versus that from H$\beta$ linewidth for NLS1s and BLS1s
in the sample of Grupe et al. (2004). The solid line in Fig. 2
means that these two masses are equal. From Fig. 2, we found that
for all BLS1s the mass from H$\beta$ width is completely larger
than that from X-ray luminosity while for NLS1s these two mass are
consistent. The distribution of $log(M(H\beta)/M(L_{\rm SX}))$ for
50 NLS1s is $0.01\pm 0.09$ with a standard deviation of 0.61. The
distribution of $log(M(H\beta)/M(L_{\rm SX}))$ for 60 BLS1s is
$1.15\pm 0.10$ with a standard deviation of 0.78. Therefore the
masses in NLS1s can be reliably derived from soft X-ray luminosity
for their super-Eddington accretion. For BLS1s it is not the case
and the soft X-ray luminosity can't derive black hole masses in
BLS1s. The consistency of $M(H\beta)$ and $M(L_{\rm SX})$ showed
that these two methods are available to estimate the mass in
NLS1s. Bian \& Zhao (2004) found that $M(H\beta)$ and $M([O~III])$
(mass from [O~III] linewidth) are not consistent for the sample of
NLS1s in SDSS (Willams 2003). Here we also found that it is the
same to the sample of Grupe et al. (2004), which is consistent
with the recent result from Grupe \& Mathur (2004). This deviation
of $M(H\beta)$ and $M([O~III])$ is nothing else than the deviation
of NLS1s from the M-$\sigma$ relation. Therefore, compared with
$M(H\beta)$ and $M(L_{\rm SX})$, $M([O~III])$ is not reliable in
NLS1s. The [O III] linewidth or $\sigma$ is not a good indicator
for black hole masses in NLS1s. This supported the idea that NLS1s
are in the early evolution stage of AGNs since the black hole is
just growing in some ways.

The uncertainties of the virial mass from the H$\beta$ linewidth
have been discussed by many authors (Krolik 2001; Wang \& Lu 2001;
Bian \& Zhao 2003b). The error in this kind of mass estimate is
about 0.5 dex (Wang \& Lu 2001). The standard deviation of
$log(M(H\beta)/M(L_{\rm SX}))$ in all 50 NLS1s is 0.61, which is
possibly from the error in the mass estimate from H$\beta$
linewidth or from $L_{\rm SX}$. The mass obtained from soft X-ray
luminosity is based on the assumption that soft hump spectrum is
$\nu F_{\nu}\propto \nu^{0}$. Moreover, the reliability in the
mass estimate from X-ray luminosity depended on whether NLS1s are
in the super-Eddington accretion process or not. There are some
NLS1s with sub-Eddington accretion (see Fig. 1). When we excluded
the NLS1s with sub-Eddington accretion, the distribution of
$log(M(H\beta)/M(L_{\rm SX}))$ is $0.18\pm 0.09$ with a standard
deviation of 0.57.

As mentioned above, we simply defined the super-Eddington
accretion as $L_{\rm bol}/L_{\rm Edd}>1$. Wang (2004) suggested a
Eddington limit relation for the super-Eddington accretion in a
hybrid structure of photon trapping and photon bubble instability.
The relation is
\begin{equation}
M_{\rm BH}=(2.9\sim 12.6)\times 10^{6}\Msolar(\frac{v_{\rm
FWHM}}{10^{3} \rm km s^{-1}})^{6.67},
\end{equation}
which is showed in Fig. 3. This relation is based on the detail
calculations of emergent spectrum from slim disk (Wang et al
1999). Thus it is independent to the estimation of black hole mass
from soft X-ray excess. Using this limit there are 16 AGNs with
super-Eddington accretion (see Table 1). Not all NLS1s satisfied
with $L_{\rm bol}/L_{\rm Edd}>1$ are satisfied with the relation
defined in above equation (5). The distribution of
$log(M(H\beta)/M(L_{\rm SX}))$ for these 16 NLS1s is $-0.61\pm
0.10$ with a standard deviation of 0.38. The masses from X-ray
luminosity for these 16 NLS1s are all systematically larger than
that derived from H$\beta$ linewidth.

\subsection{Inclinations}
The virial black hole mass derived from H$\beta$ linewidth and the
BLRs size provides a simple method to calculate the disk
inclination to the line of sight when we have other independent
methods to obtain the black hole mass (Wu \& Han 2001; Bian \&
Zhao 2002; Cao 2004; Bian 2004).

For the soft X-ray selected NLS1s, polarimetry measurements of
Grupe et al. (1998) supported a pole-on orientation (e.g. Boller
et al. 1996). However, observations of NLS1s showing strong
optical reddening/polarization and X-ray absorption suggested the
higher inclination (Goodrich 1989; Wills et al. 1992; Smith et al.
2004; Grupe et al. 2004), which is against the pole-on orientation
model of NLS1s .

From Fig. 2 and Fig. 3, we found that 16 NLS1s satisfied with the
Eddington limit relation defined by equation (5) (Wang 2004) are
all satisfied with $L_{bol}/L_{Edd}>1$ except Mrk 766. We also
noticed that the masses from X-ray luminosity for these 16 NLS1s
are all systematically larger than that derived from H$\beta$
linewidth. If we assume that the masses from X-ray luminosity in
these 16 NLS1s with super-Eddington accretion are correct, the
mass ratio between the virial mass from H$\beta$ linewidth to that
from X-ray luminosity can be approximated by $3(sin~i)^2$ (Wu \&
Han 2001; Bian \& Zhao 2002), where $i$ is the disk inclination to
the line of sight. The mean value of $log(M(H\beta)/M(L_{\rm
SX}))$ for these 16 NLS1s, $-0.61$, suggested that the mean value
of $i$ is about $17^{\circ}$, which provided new support for the
pole-on orientation effect in NLS1s.

Bian \& Zhao (2003a) suggested that the mass ratio of black hole
to the bulge in NLS1s ($M_{\rm bh}/M_{\rm bulge}$) is an order of
magnitude lower than that in BLS1s. Considering the possibility of
smaller inclination in NLS1s, we need do a careful work on their
bulges to clarify this suggestion. Bian \& Zhao (2004) used the
widths of H$\beta$ line and [O III] lines to investigate the black
hole - bulge relation in radio-loud AGNs, radio-quiet AGNs and
NLS1s. Bian \& Zhao (2004) found NLS1s deviated from the
$M-\sigma$ relation and suggested that the deviation of NLS1s may
be due to the small inclination of BLRs to the line of sight or
the reliability of the [O III] linewidth as the indicator of
stellar velocity dispersion because of its complex multiple
components. Here we found that the mean inclination is about
$17^{\circ}$ for 16 NLS1s with super-Eddington accretion defined
equation (5). The smaller inclination can partly explain the
difference of ($M_{\rm bh}/M_{\rm bulge}$) between NLS1s and
BLS1s. The smaller inclination can also partly explain why NLS1s
can't follow the $M_{\rm bh}-\sigma$ defined inactive galaxies.
Recently Wu et al. (2004) suggested a method to estimate the black
hole mass using the H$\beta$ line luminosity instead of optical
continuum luminosity. They showed that this method is better
especially for radio-loud AGNs. Although it is usually believed
that most NLS1s are radio-quiet AGNs, some cases of radio-loud
NLS1s are reported (e.g. Zhou 2003). It is necessary to use this
method to estimate the mass of NLS1s, which would be helpful to
understand the physical property of NLS1s. In order to clarify all
these questions, it is urgent to measure the bulge velocity
dispersion in NLS1s. Botte et al. (2004) recently suggested that
NLS1s follow the same $M-L_{bulge}$ relation as BLS1s, but still
deviate from $M-\sigma$ relation defined in the equation (4). Here
we should notice that the bulge luminosity given by Botte et al.
(2004) is from total host luminosity corrected by the Hubble type.
It is necessary to do a two-dimension decomposition of NLS1s host
images.

\section{Conclusion}
Different methods are used to estimate the black hole masses in a
sample of NLS1s and BLS1s (Grupe et al. 2004). The main
conclusions can be summarized as follows:
\begin{itemize}
\item{The mass from the soft X-ray bump luminosity is consistent
with that from the H$\beta$ linewidth for NLS1s. Most of NLS1s are
in the super-Eddington accretion state considering
$L_{bol}/L_{Edd}>1$. The black hole masses of NLS1s from the
H$\beta$ linewidth and from soft X-ray luminosity are reliable
while that from [O III] linewidth are not reliable.}

\item{The mean disk inclination to the line of sight in 16 NLS1s
satisfied with equation (5) is about $17^{\circ}$, which provided
new support for the pole-on orientation effect in NLS1s.}
\end{itemize}

\section*{Acknowledgments}
J.-M. Wang is greatly acknowledged for careful reading the
manuscript and the idea inputs in this paper. We thank the
anonymous referee for valuable comments. This work has been
supported by the NSFC (No. 10273007; No. 10273011) and NSF from
Jiangsu Provincial Education Department (No. 03KJB160060).

\end{document}